\documentclass[prd,reprint,showpacs,showkeys]{revtex4-1}
\usepackage{amssymb}
\usepackage{amsmath}
\usepackage{graphicx}
\usepackage[font={footnotesize,it}]{caption}

\begin{document}

\title{Thin-shell wormholes supported by total normal matter}
\author{S. Habib Mazharimousavi}
\email{habib.mazhari@emu.edu.tr}
\author{M. Halilsoy}
\email{mustafa.halilsoy@emu.edu.tr}
\affiliation{Department of Physics, Eastern Mediterranean University, Gazima\u{g}usa,
Turkey. }
\date{\today }

\begin{abstract}
The Zipoy-Voorhees-Weyl (ZVW) spacetime characterized by mass ($M$) and
oblateness ($\delta $) is proposed in the construction of viable thin-shell
wormholes (TSWs). Departure from spherical / cylindrical symmetry yields
positive total energy in spite of the fact that local energy density may
take negative values. We show that oblateness of the bumpy sources / black
holes can be incorporated as a new degree of freedom that may play role in
the resolution of the exotic matter problem in TSWs. Small velocity
perturbation reveals, however, that the resulting TSW is unstable.
\end{abstract}

\pacs{04.20.Jb, 04.20.Gz, 04.70.Bw}
\keywords{Zipoy-Voorhees-Weyl spacetime; Thin-Shell wormhole; Normal matter; 
}
\maketitle

\section{Introduction}

Until popularized by Morris and Thorne \cite{1} the idea of spacetime
wormhole introduced in the 1930s by Einstein and Rosen \cite{2} was
considered non-physical and largely as a fantasy. Although all kinds of
spherically / cylindrically symmetric metrics known to date were tried the
wormhole concept was shadowed by the required negative total energy. While
it was easy to resort to quantum field theoretical negative energy for
remedy the absence of large scale quantum systems persisted as another
serious handicap. For such reasons relying on classical physics and
searching for support within this context seems indispensable. Even to
minimize the negative (exotic) energy the idea of thin-shell wormhole (TSW)
was developed (See \cite{3} and those cited therein). By construction in all
these studies the otherwise non-traversable wormhole throat that connects
two different universes has circular topology.

In this study we add oblateness as a new degree of freedom represented by
the parameter $\delta $ ($-\infty <\delta <\infty $) and show that for
certain range of $\delta $ total energy becomes positive to avoid exotic
sources. This happens in the Zipoy-Voorhees-Weyl (ZVW) spacetime \cite{4}
with a quadrupole moment $Q=\frac{1}{3}M^{3}\delta \left( 1-\delta
^{2}\right) $, where $M=$mass of the bumpy object (or the black hole).
Naturally for $\delta =1$ one recovers the spherical Schwarzschild geometry.
It should be supplemented that integrability and chaotic behaviours of the
ZVW spacetime still is not well understood \cite{5}. Asymptotically flat,
rotating ZVW metric was discovered by Tomimatsu and Sato (TS) \cite{6},
which similar to its static predecessor remains from physics stand point yet
unclear. Once the problem of gravitational wave detection is overcome we
expect that non-Kerr (i.e., $\delta \neq 1$) multipoles of the entire TS
family can be detected. We note also that non-asymptotically flat extension
of ZVW metric is also available whose physics is yet to be understood \cite%
{7}. Herein we concentrate on the static ZVW metric in general relativity
and construct thin-shell wormhole (TSW) in this spacetime.

We should add that in the context of the modified theories of gravity,
previously, there were some attempts to introduce thin-shell wormhole
supported by positive / normal matter \cite{8}. From this token it was
realized that the normal matter is possible only for the exotic branch
solution of the Einstein-Gauss-Bonnet field equation \cite{8}.

The paper is organized as follows. Construction of TSW in ZVW spacetime is
carried out in Section II. Integration of the total energy is achieved in
Section III. Stability analysis follows in Section IV and Conclusion in
Section V completes the paper.

\section{ZVW thin-shell wormhole (TSW)}

The two parameter ZVW spacetime in the prolate spheroidal coordinates is
described by the line element%
\begin{multline}
ds^{2}=-A\left( x\right) dt^{2}+ \\
B\left( x,y\right) dx^{2}+C\left( x,y\right) dy^{2}+F\left( x,y\right)
d\varphi ^{2}
\end{multline}%
in which 
\begin{eqnarray}
A &=&\left( \frac{x-1}{x+1}\right) ^{\delta },\text{ \ }B=\frac{k^{2}}{A}%
\left( \frac{x^{2}-1}{x^{2}-y^{2}}\right) ^{\left( \delta ^{2}-1\right) } \\
C &=&B\left( \frac{x^{2}-1}{1-y^{2}}\right) ,\text{ \ }F=\frac{k^{2}}{A}%
\left( x^{2}-1\right) \left( 1-y^{2}\right) ,  \notag
\end{eqnarray}%
where $k=\frac{M}{\delta }$ and the range of coordinates are $1<x$, $0\leq
y^{2}\leq 1$, $\varphi \in \left[ 0,2\pi \right] $ and $-\infty <t<\infty $.
We note that $-\infty <\delta <\infty $ such that $\delta =0$ corresponds to
a flat spacetime and with $\delta =1$ one finds the Schwarzschild black hole
solution with the horizon located at $x=1$. For the case $\delta \neq 1$ the
hypersurface $x=1$ is a true curvature singularity (naked singularity) \cite%
{9}. As we shall see, the asymptotic behaviour of the ZV spacetime for $%
x\rightarrow \infty $ and $\delta >1$ is of our interest. It can be seen
from (1) that, in the limit $x\rightarrow \infty ,$ it becomes%
\begin{equation}
ds^{2}=-dt^{2}+k^{2}dx^{2}+k^{2}x^{2}\left( \frac{dy^{2}}{1-y^{2}}+\left(
1-y^{2}\right) d\varphi ^{2}\right)
\end{equation}%
which after a redefinition of $kx=r$ and $y=\cos \theta ,$ the line element
becomes%
\begin{equation}
ds^{2}=-dt^{2}+dr^{2}+r^{2}\left( d\theta ^{2}+\sin ^{2}\theta d\varphi
^{2}\right)
\end{equation}%
which is flat.

Construction of thin-shell wormhole (TSW) follows the standard procedure of
cutting and pasting \cite{8}. We consider two copies of ZVW spacetime and we
remove from each 
\begin{equation}
\mathcal{M}^{\pm }=\left\{ x^{\pm }<a,\text{ }1<a\right\}
\end{equation}%
in which $a=$ constant outside the singularities / horizons. We should
comment here that the minimality conditions of Hochberg and Visser \cite{10}
which is also known as \textit{the generalized flare-out conditions} in
static wormholes do not apply in the present case of TSW \cite{11}. At the
throat two spacetimes are identified to make a complete manifold. We
introduce next the induced coordinates $\xi ^{i}=\left( \tau ,y,\phi \right) 
$ on the wormhole's throat with its proper time $\tau $. The two coordinates
are related by%
\begin{equation}
g_{ij}=\frac{\partial x^{\alpha }}{\partial \xi ^{i}}\frac{\partial x^{\beta
}}{\partial \xi ^{j}}g_{\alpha \beta }
\end{equation}%
so that induced metric on the throat $\Sigma $ reads 
\begin{equation}
g_{ij}=diag\left[ -1,C\left( a\left( \tau \right) ,y\right) ,F\left( a\left(
\tau \right) ,y\right) \right] .
\end{equation}%
The Israel junction conditions \cite{12} on $\Sigma $ take the form ($c=8\pi
G=1$) 
\begin{equation}
\left\langle K_{i}^{j}\right\rangle -\left\langle K\right\rangle \delta
_{i}^{j}=-S_{i}^{j},
\end{equation}%
in which $\left\langle .\right\rangle $ stands for a jump across the
hypersurface. $K_{i}^{j}$ is the extrinsic curvature defined by 
\begin{equation}
K_{ij}^{\left( \pm \right) }=-n_{\gamma }^{\left( \pm \right) }\left( \frac{%
\partial ^{2}x^{\gamma }}{\partial \xi ^{i}\partial \xi ^{j}}+\Gamma
_{\alpha \beta }^{\gamma }\frac{\partial x^{\alpha }}{\partial \xi ^{i}}%
\frac{\partial x^{\beta }}{\partial \xi ^{j}}\right) _{\Sigma }
\end{equation}%
with the normal unit vector 
\begin{equation}
n_{\gamma }^{\left( \pm \right) }=\left( \pm \left\vert g^{\alpha \beta }%
\frac{\partial \mathcal{H}}{\partial x^{\alpha }}\frac{\partial \mathcal{H}}{%
\partial x^{\beta }}\right\vert ^{-1/2}\frac{\partial \mathcal{H}}{\partial
x^{\gamma }}\right) .
\end{equation}%
Note that $\left\langle K\right\rangle =Trace\left\langle
K_{i}^{j}\right\rangle $ and $S_{i}^{j}=$diag$\left( -\sigma ,P_{y},P_{\phi
}\right) $ is the energy momentum tensor on the thin-shell. The parametric
equation of the hypersurface $\Sigma $ is given by%
\begin{equation}
\mathcal{H}\left( x,a\left( \tau \right) \right) =x-a\left( \tau \right) =0.
\end{equation}%
The normal unit vectors to $\mathcal{M}_{\pm }$ is found to be%
\begin{equation}
n_{\gamma }^{\left( \pm \right) }=\pm \left( -\sqrt{AB}\dot{a},B\sqrt{\Delta 
},0,0\right) _{\Sigma }
\end{equation}%
with $\Delta =\frac{1}{B}+\dot{a}^{2}$ and $\dot{a}=\frac{da}{d\tau }$. The
resulting extrinsic curvature components are 
\begin{eqnarray}
K_{\tau }^{\tau \left( \pm \right) } &=&\pm \frac{\ddot{a}+\left( \frac{B_{a}%
}{B}+\frac{A_{a}}{A}\right) \frac{\dot{a}^{2}}{2}+\frac{A_{a}}{2AB}}{\sqrt{%
\Delta }} \\
K_{y}^{y\left( \pm \right) } &=&\pm \frac{C_{a}}{2C}\sqrt{\Delta },  \notag
\\
K_{\varphi }^{\varphi \left( \pm \right) } &=&\pm \frac{F_{a}}{2F}\sqrt{%
\Delta },  \notag
\end{eqnarray}%
in which a subscript $a$ stands for $\frac{\partial }{\partial a}$. The
surface energy momentum tensor has components defined as 
\begin{equation}
\sigma =-\left( \frac{C_{a}}{C}+\frac{F_{a}}{F}\right) \sqrt{\Delta }
\end{equation}%
\begin{equation}
P_{y}=\frac{2\ddot{a}+\left( \frac{B_{a}}{B}+\frac{A_{a}}{A}\right) \dot{a}%
^{2}+\frac{A_{a}}{AB}}{\sqrt{\Delta }}+\frac{\sqrt{\Delta }F_{a}}{F}
\end{equation}%
\begin{equation}
P_{\varphi }=\frac{2\ddot{a}+\left( \frac{B_{a}}{B}+\frac{A_{a}}{A}\right) 
\dot{a}^{2}+\frac{A_{a}}{AB}}{\sqrt{\Delta }}+\frac{\sqrt{\Delta }C_{a}}{C}.
\end{equation}

\begin{figure}[tbp]
\includegraphics[width=80mm,scale=0.7]{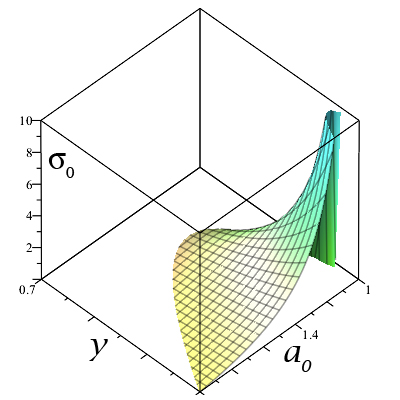} %
\captionsetup{justification=raggedright, singlelinecheck=false}
\caption{A $3D$ plot of the positive part of $\protect\sigma _{0}$ in terms
of $a_{0}$ and $y$ with $\protect\delta =2.0$. We see that although $\protect%
\sigma _{0}$ gets positive values for some interval of $y$ but it is not
positive everywhere on $y$. When the value of $\protect\delta $ decreases
the interval of $y$ on which $\protect\delta $ is positive gets smaller and
ultimately for $\protect\delta \leq 1$ the interval disappears so that $%
\protect\sigma _{0}$ gets only negative values. (Note that $\protect\sigma %
_{0}$ is an even function with respect to $y$ and only a section has been
plotted.)}
\end{figure}
\begin{figure}[tbp]
\includegraphics[width=80mm,scale=0.7]{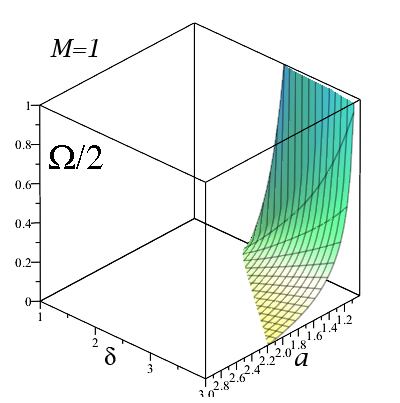} %
\captionsetup{justification=raggedright, singlelinecheck=false}
\caption{A $3D$ plot of the positive amount of energy $\frac{\Omega }{2}$
versus $a$ and $\protect\delta $ with constant mass parameter i.e. $M=1.$
This figure shows clearly that for $\protect\delta >2$ there exist $a_{c}$
in which with $1<a<a_{c}$ the total energy is positive and therefore the
resultant thin shell wormhole is supported by ordinary / normal matter. It
should be added that for a given $\protect\delta >2$, there exists $%
1<a<a_{c} $ which leads to a physically accepted TSW.}
\end{figure}

\begin{figure}[tbp]
\includegraphics[width=80mm,scale=0.7]{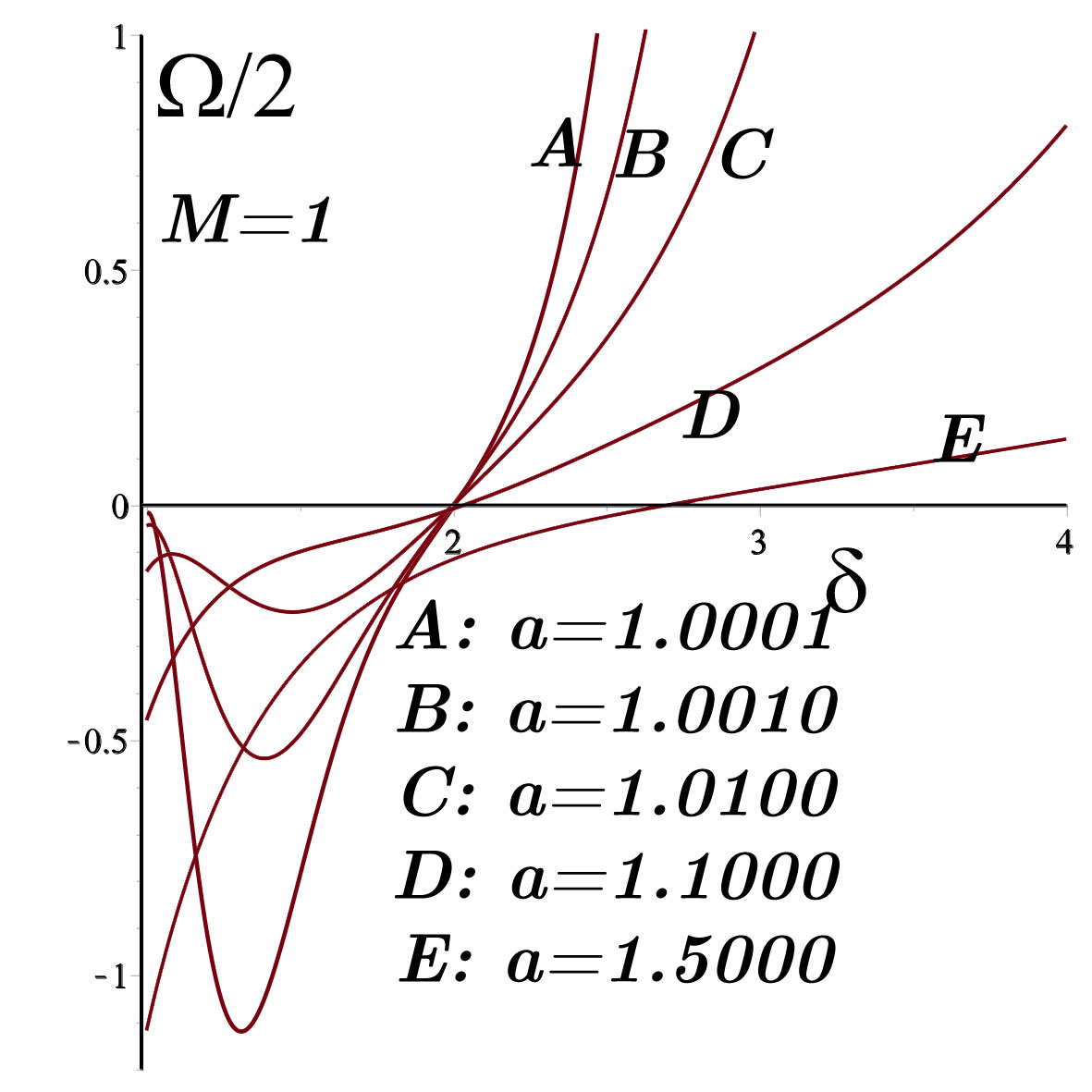} %
\captionsetup{justification=raggedright, singlelinecheck=false}
\caption{$\frac{\Omega }{2}$ versus $\protect\delta $ for different values
of $a=1.0001,1.0010,1.0100,1.1000,1.5000$ and $M=1.$ As we commented in Fig.
2, here it is more clear that when the value of $a\rightarrow 1$, the
threshold $\protect\delta $ which admits a positive total energy approaches
to $\protect\delta =2.$ This means that $\protect\delta =2$ is a critical
value to have a thin shell wormhole supported by ordinary matter. For larger 
$a$ the threshold $\protect\delta $ gets larger values.}
\end{figure}

\section{Positive Matter Sources}

The energy-momentum components at equilibrium condition, i.e. $a=a_{0}=$%
constant with $\dot{a}=\ddot{a}=0$ yields%
\begin{eqnarray}
\sigma _{0} &=&\left. -\left( \frac{C_{a}}{C}+\frac{F_{a}}{F}\right) \frac{1%
}{\sqrt{B}}\right\vert _{a=a_{0}}  \notag \\
P_{y0} &=&\left. \left( \frac{A_{a}}{A}+\frac{F_{a}}{F}\right) \frac{1}{%
\sqrt{B}}\right\vert _{a=a_{0}}, \\
P_{\phi 0} &=&\left. \left( \frac{A_{a}}{A}+\frac{C_{a}}{C}\right) \frac{1}{%
\sqrt{B}}\right\vert _{a=a_{0}}.  \notag
\end{eqnarray}%
The explicit form of the energy momentum components are then found to be%
\begin{eqnarray}
\sigma _{0} &=&\frac{2\left[ 2\delta \left( a_{0}^{2}-y^{2}\right)
-2a_{0}^{3}+a_{0}\left( 1+y^{2}\right) +a_{0}\delta ^{2}\left(
y^{2}-1\right) \right] }{\left( a_{0}^{2}-1\right) \left(
a_{0}^{2}-y^{2}\right) },  \notag \\
P_{y0} &=&\frac{2a_{0}}{a_{0}^{2}-1}, \\
P_{\phi 0} &=&\frac{2a_{0}\left[ a_{0}^{2}-1+\left( 1-y^{2}\right) \delta
^{2}\right] }{\left( a_{0}^{2}-1\right) \left( a_{0}^{2}-y^{2}\right) }. 
\notag
\end{eqnarray}%
Note that since $1<a_{0}$ there are no singularities in the foregoing
expressions. From the conditions $a_{0}>1$ and $y<1$ one observes that $%
P_{y0}$ and $P_{\phi 0}$ are both positive while $\sigma _{0}$ may be
positive, negative or zero. Fig. 1 displays $\sigma _{0}$ in terms of $a_{0}$
and $y$. We observe that for $\delta >1$ there exists regions that $\sigma
_{0}$ becomes positive. This is seen clearly in Fig. 1. Also in the interval
on which $\sigma _{0}\geq 0$ the weak and strong energy conditions are
satisfied.

We note that, in Fig. 1 the energy density $\sigma _{0}$ is shown in terms
of $y$ and $a_{0}$ but only $y$ is variable and $a_{0}$ is fixed for a
specific TSW. This means that at the throat $x=a_{0}$ and only $y$ and $\phi 
$ are variable. As one sees from (18), there is angular symmetry and as a
result, not only $\sigma _{0}$ but also $P_{y0}$ and $P_{\phi 0}$ are only
functions of $y.$ In addition, $P_{y0}$ is a constant function of $y$.
Therefore once we set the radius of the throat i.e., $a_{0},$ our energy
momentum tensor's components are left with the only variable $y.$ Hence,
depending on $y$, the energy density of the TSW i.e. $\sigma _{0}$ is
locally positive or negative. This is what we see in Fig. 1. Of course, the
situation is completely different for $P_{y0}$ and $P_{\phi 0}$ which are
positive for entire domain of $y.$

In addition to the energy conditions we are mainly interested in the total
energy supporting the TSW given by 
\begin{equation}
\Omega =2\int\nolimits_{0}^{2\pi
}\int\nolimits_{0}^{1}\int\nolimits_{1}^{\infty }\sigma _{0}\mathcal{\delta }%
\left( x-a_{0}\right) \sqrt{-g}dxdyd\phi ,
\end{equation}%
which simplifies to 
\begin{equation}
\Omega =4\pi \int\nolimits_{0}^{1}\sigma _{0}\sqrt{-g_{0}}dy.
\end{equation}%
In Eq. (19) $\mathrm{\delta }\left( x-a_{0}\right) $ is the Dirac delta
function. In Fig. 2 we plot $\Omega $ versus $a$ and $\delta $ with fixed
value of mass $M=1.$ Fig. 3 reveals more details. These plots overall show
that TSW\ supported by normal matter is possible provided $\delta >2$. That
explains also why given ordinary matter alone in Schwarzschild spacetime
with $\delta =1,$ there was no such traversable wormhole. In this regard let
us add that even an arbitrarily small energy condition violation is
considered respectable \cite{13}.

\section{Stability Analysis}

In this section we apply the small velocity perturbation on the shell under
the condition that the EoS of the TSW after the perturbation is same as its
EoS at its static equilibrium. This is possible if the perturbation process
occurs slow enough in which all the intermediate states can also be
considered as static equilibrium points. Therefore the EoS after the
perturbation reads%
\begin{equation}
\frac{P_{y}}{\sigma }=-\frac{\frac{A_{a}}{A}+\frac{F_{a}}{F}}{\frac{C_{a}}{C}%
+\frac{F_{a}}{F}}
\end{equation}%
and%
\begin{equation}
\frac{P_{\varphi }}{\sigma }=-\frac{\frac{A_{a}}{A}+\frac{C_{a}}{C}}{\frac{%
C_{a}}{C}+\frac{F_{a}}{F}}.
\end{equation}%
These in explicit form amount to%
\begin{equation}
2\ddot{a}+\left( \frac{B_{a}}{B}\right) \dot{a}^{2}=0
\end{equation}%
which upon integration yields%
\begin{equation}
\dot{a}=\dot{a}_{0}\sqrt{\frac{B_{0}}{B}}=\dot{a}_{0}\sqrt{\frac{\left( 
\frac{a_{0}+1}{a_{0}-1}\right) ^{\delta }\left( \frac{a_{0}^{2}-1}{%
a_{0}^{2}-y^{2}}\right) ^{\left( \delta ^{2}-1\right) }}{\left( \frac{a+1}{%
a-1}\right) ^{\delta }\left( \frac{a^{2}-1}{a^{2}-y^{2}}\right) ^{\left(
\delta ^{2}-1\right) }}}.\text{ \ }
\end{equation}%
In Figs. 4 and 5 we plot $\dot{a}$ in terms of $a$ and $y$ for the initial
values $a_{0}=1.2$ and $\dot{a}_{0}=\pm 0.1,$ respectively, and $\delta =2.5$%
. As one can see in both cases the velocity does not get zero, which means
that the throat does not go back to its initial position.

\begin{figure}[tbp]
\includegraphics[width=80mm,scale=0.7]{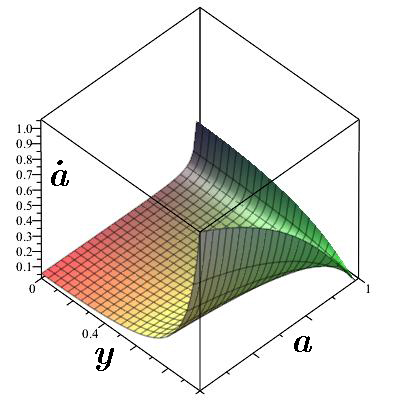} %
\captionsetup{justification=raggedright, singlelinecheck=false}
\caption{$\dot{a}$ versus $a$ and $y$ with $\protect\delta =2.5,$ $\dot{a}%
_{0}=0.1$ and $a_{0}=1.2.$ The velocity never vanishes which is an
indication of instability of the throat under the small velocity
perturbation.}
\end{figure}
\begin{figure}[tbp]
\includegraphics[width=80mm,scale=0.7]{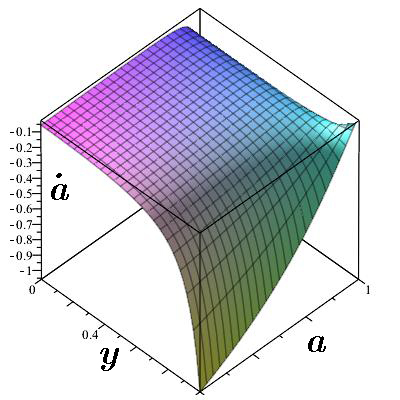} %
\captionsetup{justification=raggedright, singlelinecheck=false}
\caption{A $3D$ plot of $\dot{a}$ versus $a$ and $y$ with $\protect\delta %
=2.5,$ $\dot{a}_{0}=-0.1$ and $a_{0}=1.2.$ The velocity fails to vanish
which amounts to instability of the throat under the small velocity
perturbation.}
\end{figure}

\section{Conclusion}

We considered the possibility of having physical wormhole solution in
Einstein's general relativity which is supported by normal matter and the
energy conditions are satisfied. Our emphasis is on Einstein's general
relativity instead the modified theories such as the Lovelock theory. In
such theories there were attempts to find such physical wormholes \cite{8}
with non-physical solution and it was shown that a TSW supported by positive
energy was possible. In this context we have shown that TSWs with oblate
sources can be employed to admit an overall physical (i.e. non-exotic)
matter even in Einstein's general relativity. As we have depicted in
figures, for $\delta >2$ not only the total energy of the wormhole is
positive but also the WECs and SECs are satisfied for limited $y$-interval,
which increases for larger $\delta .$ Certain range of deviation from
spherical symmetry can be chosen from the total energy integral to render
this possible. Locally it can easily be checked from Eq. (16) for $y=0$, for
instance, that we have a negative energy density, however, this is
compensated within the total energy expression. It is also expected that
once the metric and the surface energy-momentum become time dependent energy
conservation on the thin shell will not be valid any more. Another important
aspect concerning TSWs which has not been discussed here is their stability
against perturbations. It is shown that small velocity perturbations in the $%
x-$direction leads to an unstable wormhole throat. Finally, from curiosity
we wish to ask: \textit{does the deformation parameter }$\delta $\textit{\
saves wormholes other than TSWs in Einstein's theory?}. It remains to be
seen.

\bigskip


\begin{thebibliography}{99}
\bibitem{1} M. S. Morris and K. S. Thorne, Am. J. Phys. \textbf{56}, 5
(1998).

\bibitem{2} A. Einstein and and N. Rosen, Phys. Rev. \textbf{48}, 73 (1935).

\bibitem{3} M. Visser, \textit{Lorentzian Wormholes: From Einstein to
Hawking (American Institute of Physics, New York, 1995)};

M. Visser, Phys. Rev. D \textbf{39}, 3182 (1989);

M. Visser, Nucl. Phys. \textbf{B} 328, 203 (1989);

E. Poisson and M. Visser, Phys. Rev. D \textbf{52}, 7318 (1995);

P. R. Brady, J. Louko and E. Poisson, Phys. Rev. D \textbf{44}, 1891 (1991);

M. Ishak and K. Lake, Phys. Rev. D \textbf{65}, 044011 (2002);

E. F. Eiroa and C. Simeone, Phys. Rev. D \textbf{70}, 044008 (2004);

E. F. Eiroa and C. Simeone, Phys. Rev. D \textbf{81}, 084022 (2010);

E. F. Eiroa and C. Simeone, Phys. Rev. D \textbf{71}, 127501 (2005);

E. F. Eiroa, Phys. Rev. D \textbf{78}, 024018 (2008);

E. F. Eiroa and G. F. Aguirre, Eur. Phys. J. C \textbf{72,} 2240 (2012);

C. Bejarano and E. F. Eiroa, Phys. Rev. D \textbf{84}, 064043 (2011);

E. F. Eiroa and C. Simeone, Phys. Rev. D \textbf{76,} 024021 (2007);

E. F. Eiroa, Phys. Rev. D \textbf{80,} 044033 (2009);

M. G. Richarte, Phys. Rev. D \textbf{82}, 044021 (2010);

E. F. Eiroa and C. Simeone Phys. Rev. D \textbf{82}, 084039 (2010);

F. S. N. Lobo, Phys. Rev. D \textbf{71}, 124022 (2005);

N. M. Garcia, F. S. N. Lobo and M. Visser, Phys. Rev. D \textbf{86}, 044026
(2012);

J. P. S. Lemos and F. S. N. Lobo, Phys. Rev. D \textbf{78}, 044030 (2008);

M. H. Dehghani and M. R. Mehdizadeh, Phys. Rev. D \textbf{85}, 024024 (2012);

M. Sharif and M. Azam, JCAP \textbf{04}, 023 (2013);

M. Sharif and M. Azam, JCAP \textbf{05}, 25 (2013);

M. Sharif and M. Azam, Eur. Phys. J. C \textbf{73,} 2407 (2013);

M. Sharif and M. Azam, Eur. Phys. J. C \textbf{73,} 2554 (2013);

M. Jamil, M. U. Farooq and M. A. Rashid, Eur. Phys. J. C \textbf{59,} 907
(2009).

\bibitem{4} H. Weyl, Ann. Physik, \textbf{54}, 117 (1917);

D. \thinspace M. Zipoy, J. Math. Phys. (N.Y.) \textbf{7}, 1137 (1966);

B. \thinspace H. Voorhees, Phys. Rev. D \textbf{2}, 2119 (1970).

\bibitem{5} G. L.-Gerakopoulos, Phys. Rev. D \textbf{86}, 044013 (2012);

N. A. Collins and S. A. Hughes, Phys. Rev. D \textbf{69}, 124022 (2004).

\bibitem{6} A. Tomimatsu and H. Sato, Phys. Rev. Lett. \textbf{29}, 1344
(1972);

A. Tomimatsu and H. Sato, Prog. Theor. Phys. \textbf{50}, 95 (1973).

\bibitem{7} R. M. Kerns and W. J. Wild, Phys. Rev. D \textbf{26}, 3726
(1982);

M. Halilsoy, J. Math. Phys. \textbf{33}, 4225 (1992).

\bibitem{8} M. G. Richarte and C. Simeone, Phys. Rev. D \textbf{76}, 087502
(2007); Phys. Rev. D \textbf{77}, 089903(E) (2008);

C. Simeone, Phys. Rev. D \textbf{83}, 087503 (2011);

S. H. Mazharimousavi, M. Halilsoy and Z. Amirabi, Phys. Rev. D \textbf{81},
104002 (2010);

S. H. Mazharimousavi, M. Halilsoy and Z. Amirabi, Class. Quantum Grav. 
\textbf{28}, 025004 (2011);

T. Bandyophyay and S. Chakraborty, Class. Quantum Grav. \textbf{26,} 085005
(2009);

P. Kanti, B. Kleihaus and J. Kunz, Phys. Rev. Lett. \textbf{107}, 271101
(2011);

P. Kanti, B. Kleihaus and J. Kunz, Phys. Rev. D \textbf{85}, 044007 (2012).

\bibitem{9} O. Obreg\'{o}n, H. Quevedo and M. P. Ryan, JHEP \textbf{07}, 005
(2004);

O. Obreg\'{o}n, H. Quevedo and M. P. Ryan, Phys. Rev. D \textbf{70}, 064035
(2004);

L. Herrera, F. M. Paiva, and N. O. Santos, J. Math. Phys. (N.Y.) \textbf{40}%
, 4064 (1999);

H. Kodama andW. Hikida, Classical Quantum Gravity \textbf{20}, 5121 (2003);

D. Papadopoulos, B. Stewart, and L. Witten, Phys. Rev. D \textbf{24}, 320
(1981).

\bibitem{10} D. Hochberg and M. Visser, Phys. Rev. D \textbf{56},4745 (1997).

\bibitem{11} S. H. Mazharimousavi and M. Halilsoy, (2014) "\textit{Flare-out
conditions in static thin-shell wormholes}" arXiv:1311.6697.

\bibitem{12} W. Israel, Nuovo Cimento \textbf{44B}, 1 (1966);

V. de la Cruzand W. Israel, Nuovo Cimento \textbf{51A}, 774 (1967);

J. E. Chase, Nuovo Cimento \textbf{67B}, 136. (1970);

S. K. Blau, E. I. Guendelman, and A. H. Guth, Phys. Rev. D \textbf{35}, 1747
(1987);

R. Balbinot and E. Poisson, Phys. Rev. D \textbf{41}, 395 (1990).

\bibitem{13} M. Visser, S. Kar and N. Dadhich, Phys. Rev. Lett. \textbf{90},
201102 (2003).
\end{thebibliography}
\end{document}